\documentclass[aps,prl,twocolumn,groupedaddress]{revtex4}

\usepackage{graphicx}
\usepackage{graphics}
\usepackage{dcolumn}
\usepackage{bm}
\usepackage{color}

\def\dfrac#1#2{{\displaystyle\frac{#1}{#2}}}

\begin{document}

\preprint{APS/123-QED}

\title{Ferroelectric Soft Mode in Pb(Mg$_{1/3}$Nb$_{2/3}$)O$_3$}

\author{Hiroki Taniguchi}
 \email{taniguchi.h.aa@m.titech.ac,jp}
\affiliation{Materials and Structure Laboratory, Tokyo Institute of Technology, Yokohama 226-8503, Japan}

\author{Masanori Oishi}
\affiliation{Materials and Structure Laboratory, Tokyo Institute of Technology, Yokohama 226-8503, Japan}

\author{Desheng Fu}
\affiliation{Division of Global Research Leaders, Shizuoka University, Hamamatsu 432-8561, Japan}

\author{Mitsuru Itoh}
\affiliation{Materials and Structure Laboratory, Tokyo Institute of Technology, Yokohama 226-8503, Japan}

\date{\today}


\begin{abstract}
 Ferroelectric soft mode in Pb(Mg$_{1/3}$Nb$_{2/3}$)O$_3$ (PMN) has been clearly resolved by precision Raman scattering measurements for the first time. A polarization direction of the incident laser was chosen along [110] in cubic configuration in order to eliminate strong scattering components around 45 cm$^{-1}$, which always smeared the low-frequency spectra of PMN. The soft mode frequency $\omega_{\scriptsize{\mbox{s}}}$ ($=\sqrt{\omega_0^2-\gamma^2}$) softens around 200 K, where $\omega_0$ and $\gamma$ are a harmonic frequency and a damping constant, respectively. The present result evidenced that the origin of the polarizationthe in PMN is the soft mode.
\end{abstract}

\maketitle

	Pb(Mg$_{1/3}$Nb$_{2/3}$)O$_3$ (PMN) has been recognized as a prototypical compound among relaxors, which belong to a class of materials with giant and dispersive dielectric response due to nanoscopic heterogeneity in the matrix.\cite{Smolensky2, Cross, Samara4} Elucidating the nature of relaxor behavior is significant for a development of electromechanical/optic devices, and a fundamental understanding of phase transition properties in disordered materials.

	It has been known that PMN has two characteristic temperatures: the one is the Burns temperature $T_d \sim$ 620 K\cite{Burns5}, at which nanoscopic polar clusters appear, and the other is $T_0 \sim$ 210 K where a macroscopic ferroelectric phase can be induced by an electric bias field larger than 2 kV/cm.\cite{Ye2} In contrast to the conventional understanding that relaxors do not undergo macroscopic ferroelectric ordering without the electric field, recent investigations have clarified in PMN certain anomalies suggesting, at a microscopic scale at least, some kind of ferroelectric phase transition at $T_0$ even under the zero bias field. NMR study has evidenced that an additional Pb shift along a rhombohedral [111] direction suddenly occurs at $T_0$, further to the spherical-shell-type displacement in the high-temperature region.\cite{Blinc2, Vakhrushev} Barkhausen jump observed by optical studies, on the other hand, strongly indicates that the local symmetry lowering in the low temperature region is due to an appearance of microscopic ferroelectric domains rather than development of dipoles in glassy state.\cite{Westphal} Considering form these results, it seems plausible that PMN is incipient ferroelectric and undergoes \textit{a ferroelectric state broken up into nanodomains under the constraint of quenched random field} (or sometimes it is called as \textit{a frustrated phase transition}).\cite{Blinc2} However, a clear physical picture has been still veiled.

	An important problem to be solved is a driving interaction for the local polarization in PMN. In order to clarify this point from a viewpoint of lattice dynamics, a number of spectroscopic studies have been performed with, for instance, a neutron inelastic scattering\cite{Gehring, Wakimoto}, an infrared absorption\cite{Kampa2} and so on. All of them suggested an existence of a ferroelectric soft mode that drives the polarization in PMN. However, the accurate observation of the critical behavior has been still lacking due to the experimental difficulty on low-frequency spectroscopies of relaxors.

	Here we show the precise soft mode dynamics in PMN observed by rigorous Raman scattering experiments. It should be noted here that previous light scattering studies chose [100] cubic axis for the polarization direction of the incident laser.\cite{Siny, Ohwa, Svitelskiy} In this scattering configuration, however, strong F$_{2g}$ component at 45 cm$^{-1}$, which stems from Mg : Nb = 1 : 1 chemically ordered $Fm\bar{3}m$ region, always smeared the low frequency part of the spectra. In the present study, we found from careful checking of an angular dependence of Raman spectra that the strong F$_{2g}$ component can be eliminated with the polarization direction along [110] axis and the crossed nicols configuration. In this treatment we have successfully resolved the precise temperature dependence of the soft mode in PMN for the first time. The observed soft mode softens at around 200 K with the conventional manner. The heavily damped feature of the soft mode suggested strongly perturbed soft phonon dynamics by the disordered matrix of PMN.

	PMN single crystals used in the present study were synthesized by the conventional columbite method, and the surfaces were polished into optical quality. A 488 nm Ar$^+$ laser was employed as the incident laser with a power of 10 mW on sample. The scattered light was collected with backscattering geometry, and analyzed by Jovin Yvon triple-monochromator T64000. The deformation of low-frequency spectral edge due to an optical slit, which sometimes obscures the low-frequency spectra, was carefully eliminated by rigorous optical alignments. Temperature of the samples was controlled by Linkam THMS600 and Oxford microstat with the stability of $<$ $\pm$1.0 K and $<$ $\pm$0.1 K, respectively.

\begin{figure}[t]
\includegraphics[width=\linewidth]{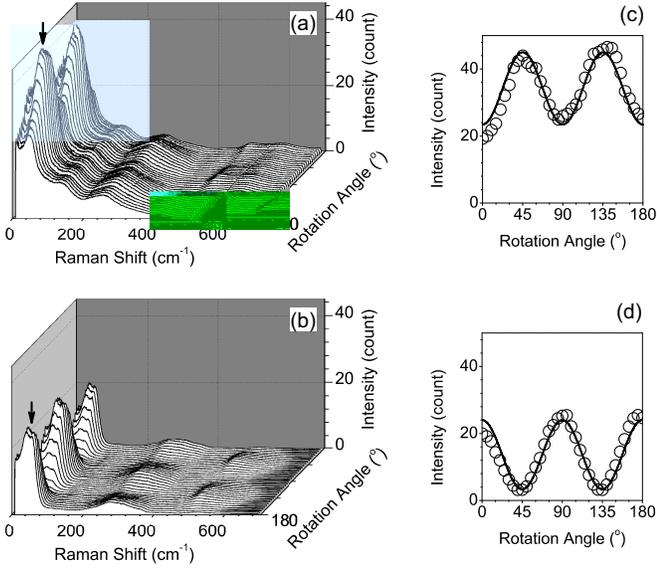}
\caption{Angular dependences of Raman spectra of PMN observed at room temperature. The spectra were observed with both (a) parallel and (b) crossed nicols configurations. Arrows in panels (a) and (b) indicate the strong F$_{2g}$ components stem from $Fm\bar{3}m$ symmetry due to the 1 : 1 chemically ordered regions. Solid lines in panels (c) and (d) present calculated angular dependences of the F$_{2g}$ mode intensities for the parallel and crossed nicols configurations, respectively. Open circles in the panels are the observed ones.}
\label{fig:angle}
\end{figure}
	In order to determine which scattering configuration is most suitable to resolve the low-frequency Raman spectra in PMN, we first checked angular dependences of Raman spectra at room temperature. Figures \ref{fig:angle} (a) and (b) are the angular dependences of Raman spectra observed by parallel and crossed nicols configurations, respectively. Sample rotations run from $z(x,x)\bar{z}$ to $z(\bar{x},\bar{x})\bar{z}$ via $z(y,y)\bar{z}$ for the parallel nicols, and from $z(x,y)\bar{z}$ to $z(\bar{x},\bar{y})\bar{z}$ via $z(xy,\bar{x}y)\bar{z}$ for the crossed one. As clearly seen in the figures, the strong F$_{2g}$ components pointed by arrows periodically change their intensities. The important result here is that, in the crossed nicols configuration [Fig. \ref{fig:angle} (b)], the F$_{2g}$ component vanishes when the scattering configration is rotated from the cubic axis by 45 degrees. Open circles in Figs. \ref{fig:angle} (c) and (d) denote angular dependences of intensities of F$_{2g}$ components at 45 cm$^{-1}$ for the parallel and crossed nicols configurations, respectively. The solid lines were calculated from a coordinate rotation of the Raman tensor. The results indicate again that if we choose the [110] cubic axis for the polarization direction of the incident laser and adopt the crossed nicols configuration, we can observe the low frequency spectra of PMN without any disturbance by the strong F$_{2g}$ component.

\begin{figure}[t]
\includegraphics[width=\linewidth]{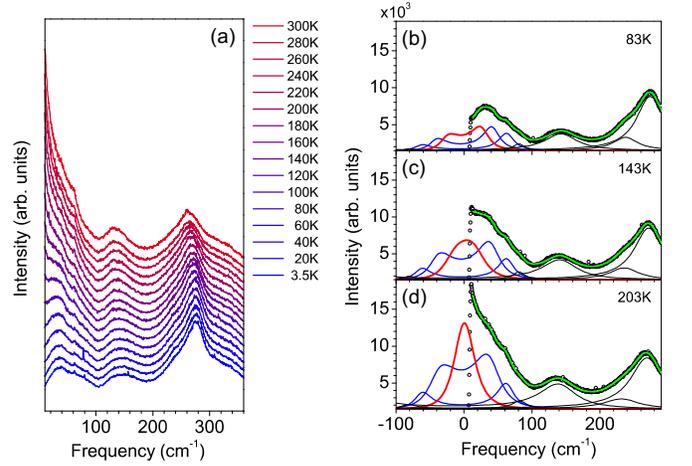}
\caption{(a) Raman spectra of PMN observed at several temperatures ranging from 3.5 K to 300 K. A Base line for each spectra are systematically shifted. Panels (b), (c), and (d) show examples of the fit for the spectra observed at 83 K, 143 K, and 203 K, respectively. See text for details.}
\label{fig:TempDep}
\end{figure}

	A temperature dependence of the low frequency Raman spectra observed in the $z(xy,\bar{x}y)\bar{z}$ scattering geometry is shown in Fig. \ref{fig:TempDep} (a) with the temperature range from 3.5 K to 300 K. As seen in the figure clearly, a dramatic change of the spectral profile has been found in the low frequency region. The central-component-like spectral feature at the high temperature region gradually transforms into a Raman side band and hardens with decreasing temperature. This result obviously indicates a soft-mode-type critical dynamics as an origin of the frustrated phase transition at $T_0$. It should be stressed here that this mode was observed from the matrix of PMN single crystal, because we eliminated the contribution from the chemically ordered $Fm\bar{3}m$ region by choosing the appropriate scattering configuration. Therefore, the observed soft mode behavior directly manifests the phase transition dynamics of the matrix area.

	Spectral analyses have been performed in order to obtain a quantitative temperature dependence of the soft mode, with a combination of damped harmonic oscillator functions,
\begin{eqnarray}
I(\omega) = F(\omega,T)\cdot\sum_i\dfrac{2A_i\omega_{0i}^2\gamma_i\omega}{(\omega_i^2 - \omega^2)^2 + 4\gamma_i^2\omega^2}
\label{eq:DHO}
\end{eqnarray}
, where $A$, $\omega_0$, and $\gamma$ denotes an intensity, a harmonic frequency and a damping constant of corresponding excitations indicated by subscripts, respectively. The temperature factor $F(\omega,T)$ is given by following equations;
\begin{eqnarray}
F(\omega,T) = \left\{
\begin{array}{cc}
n(\omega)+1 & \mbox{(Stakes part)} \\
n(\omega) & \mbox{(anti-Stokes part)} 
\end{array}\right. \\
n(\omega) = \left[ \exp\left( \dfrac{\hbar\omega}{k_BT} \right)-1 \right]^{-1}. \quad\quad\quad
\end{eqnarray}

	Figures \ref{fig:TempDep} (b), (c), and (d) present examples of the fitting for the spectra observed at 83 K, 143 K, and 203 K, respectively. Red lines in the figures indicate the soft modes, which were decomposed by calculations. As seen in the figures, the soft mode spectrum shows a distinct underdamped nature at 83 K, and with increasing temperature it softens toward the low frequency side. Finally, the soft mode becomes overdamped, showing the central-peak-like profile at 203 K. Peaks indicated by blue lines are caused by the leakage of the strong components, which are observed in $z(x,y)\bar{z}$ scattering geometry. They increase the intensities without shifting the frequencies as elevating temperature, being in excellent agreement with the temperature dependence of the strong F$_{2g}$ components, which were observed in the previous Raman scattering study.\cite{Svitelskiy} The other components reproduced by black lines show little temperature dependences, and have almost no effect on the spectral fitting of the soft mode. Finally, the calculations gave systematic and  satisfactory fits for all the spectra observed in the present study as presented by green lines.

\begin{figure}[t]
\includegraphics[width=\linewidth]{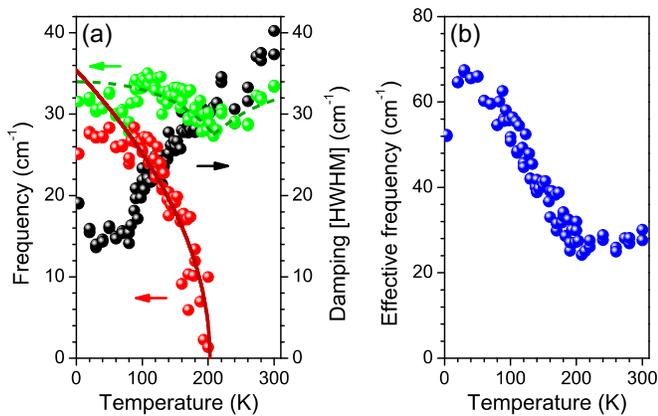}
\caption{Green and black circles in the Panel (a) denote the temperature dependence of the harmonic frequency $\omega_0$ and the damping constant $\gamma$ of the soft mode obtained by the fitting with Eq. (\ref{eq:DHO}). The soft mode frequency $\omega_{\scriptsize{\mbox{s}}}$ indicated by red circles in panel (a) were calculated by the equation $\omega_{\scriptsize{\mbox{s}}} = \sqrt{\omega_0^2-\gamma^2}$. A Curie-Weiss fit for $\omega_{\scriptsize{\mbox{s}}}$ is presented by a solid line. Blue circles in the panel (b) depicts the temperature dependence of the effective frequency, which was calculated by $\omega_0^2/\gamma$.}
\label{fig:softmode}
\end{figure}
	The temperature dependence of the soft mode frequency $\omega_{\scriptsize{\mbox{s}}}$, the harmonic frequency $\omega_0$, and the damping constant $\gamma$ that were obtained by the fit with eq. (\ref{eq:DHO}) are presented in Fig. \ref{fig:softmode} (a) with red, green and black circles, respectively, where $\omega_{\scriptsize{\mbox{s}}}$ was calculated from $\omega_0$ and $\gamma$ with the relation, $\omega_{\scriptsize{\mbox{s}}} = \sqrt{\omega_0^2-\gamma^2}$. A broken line in the figure is an eye-guide for the harmonic frequency. Surprisingly, the soft mode in PMN softens like a conventional manner as usual displacive-type ferroelectric compounds do. This result clearly evidences that intrinsically the matrix has a ferroelectric instability due to the soft mode. Exactly speaking, the ferroelectric ordering is definitely taken place in the certain region where the soft mode softens at least, in spite of the fact that PMN does not undergo the macroscopic ferroelectric phase. The temperature dependence of the soft mode was tentatively examined by a conventional Curie-Weiss law, $\omega_{\scriptsize{\mbox{s}}} = C|T-T_0|^{1/2}$. The best fit curve was obtained with $C = 2.5$ and $T_0 =202.6$ K, as indicated by a solid line in the figure.

	Even in the high temperature region above $T_c$, the soft mode spectra remains observable, though it is nominally Raman inactive. It is probably due to the local non-centrosymmetric distortion, which is closely related to the onset of polar nano regions at $T_d$. Above $T_0$ as shown in the Figs \ref{fig:TempDep} (a) and \ref{fig:softmode} (a), the soft mode is completely overdamped. In order to obtain an overall picture of the soft mode, we analyzed the soft mode dynamics by the effective frequency $\omega_0^2/\gamma$, which corresponds to the frequency of loss maximum. The result is presented in Fig. \ref{fig:softmode} (b). As seen in the figure, there is an apparent kink around 200 K, confirming certain anomaly in the soft mode dynamics at $T_0$.

	The previous results for the low-frequency spectroscopy on PMN have been in controversial situation. Although it is not our focus to compare the present result with previous ones one by one, it seems important to discuss briefly consistency of the present result with them. As a counterpart of the present soft mode, the FTIR transmission study reported the low frequency absorption band below 30 cm$^{-1}$, however it was assigned as a relaxational excitation of polar clusters from the central-peak-like spectral feature.\cite{Kampa2} This apparent discrepancy is probably caused by the difference of the frequency resolution. The present Raman measurement has several times higher frequency resolution than the FTIR study, and in general the heavily damped low frequency excitation is quite difficult to resolve by the FTIR measurement. Therefore, it is reasonable to assign the low frequency excitation to the soft mode as indicated by the present study, and this conclusion does not contradict with the observed FTIR spectra. Neutron diffraction studies, on the other hand, have reported the soft mode is located at 90 cm$^{-1}$ near zero Kelvin, and softens as increasing temperature.\cite{Wakimoto} This mode was first proposed as an evidence of ferroelectric ordering in PMN, and the similar mode was also detected in the FTIR study. However this mode was not observed by the present study. This controversy is resolved if we accept the phase transition from cubic $Pm\bar{3}m$ to rhombohedral $R3m$, which has been implied by many studies, at least in the region where the soft mode softens. In this phase transition sequence, the soft mode is considered to split from the triply degenerated $F_{1u}$ mode to the $A_1$ mode and the doubly degenerated $E$ mode. Generally in ferroelectric compound, the $E$ mode is smaller in frequency than $A_1$ mode. Therefore, the soft modes, which are observed by the present study and neutron diffraction, can be assigned to the the $E$ and $A_1$ modes, respectively. Raman selection rule is also consistent with this assignment. In the present study, we measured the spectra with the crossed nicols configuration in order to eliminate the strong $F_{2g}$ component. With this configuration we can only detect the $E$ mode, and $A$ mode appears only in the parallel nicols one. Unfortunately, the strong $F_{2g}$ component completely smears the low frequency spectra  in the parallel nicols configuration, therefore we can not access the $A_1$ mode by the Raman scattering. The physical picture of PMN indicated by this speculation is that the rhombohedral ferroelectric ordering is, at least locally, induced by the softening of $F_{1u}$ soft mode at $T_0$ in the matrix, which is separated from the temperature independent 1 : 1 chemically ordered $Fm\bar{3}m$ regions.

\begin{figure}[t]
\includegraphics[width=.85\linewidth]{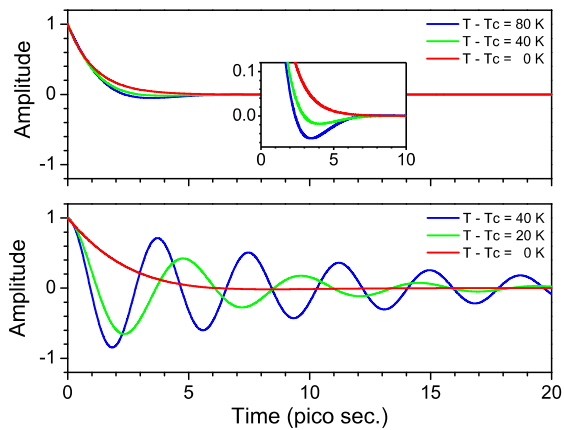}
\caption{Panels (a) and (b) shows the softening of the soft mode in PMN and CdTiO$_3$ with a time domain expression, respectively. The inset in panel (a) presents the magnified picture of the soft mode in PMN for the time interval from 0 to 10 pico sec. Curves are calculated by the parameters obtained by the fitting of Raman spectra. The Raman spectra of the soft mode in CdTiO$_3$ was previously reported in Ref. \cite{Taniguchi5}.}
\label{fig:oscillation}
\end{figure}
	Finally, but it must by stressed that the soft mode in PMN has pretty different feature in the critical dynamics from that of conventional displacive-type compounds. Figures \ref{fig:oscillation} (a) and (b) depict temperature dependences of the soft mode oscillations in PMN and that in the ideal displacive-type ferroelectric material CdTiO$_3$\cite{Taniguchi5}, respectively. In order to emphasize the difference between them, the oscillations are visualized by a time domain expression, where the all curves were calculated from the parameters obtained by the fits of the observed Raman spectra. As can be seen in the figure, the soft mode in PMN is governed rather by a damping than by a decrease of the harmonic frequency, in contrast to the typical displacive-type phase transition as indicated by the case of CdTiO$_3$ in Fig. \ref{fig:oscillation} (b). The heavily damped soft mode dynamics in PMN is provably caused by the disturbance of long range oscillation due to the chemically ordered regions, which are randomly dispersed in the matrix. in spite of the heavily damped or in other words \textit{relaxational-mode-like} characteristic of the soft mode, it is important that the polarization in PMN is driven by the \textit{lattce vibration} as evidenced by the critical softening of the underdamped soft mode. Therefore, the development of polar nano regions at $T_d$ and the dispersive dielectric anomaly in relaxors would be result in a problem of anharmonic lattice dynamics in the nanometric disordered media, which involves ferroelectric instability.

In conclusion, we have observed for the first time the precise temperature dependence of the low frequency soft mode in PMN, after optimizing the best scattering configuration. The observed soft mode softens with underdamped oscillation at around $T_0$ where the electric field induced ferroelectric phase transition has been known to occur. Results strongly indicate that the ferroelectric ordering is taken place even under the zero bias field at least in the region where the soft mode softens. It can be also concluded from the result that the origin of the polarization in PMN is the soft mode displacement. The heavily damped characteristic of the soft mode oscillation, in spite of the underdamped nature, reflects the nanoscopic disorder in the matrix of PMN. Provably, the critical soft mode dynamics in the heavily disordered matrix is the key to resolve the mechanism of relaxor properties.


\end{document}